\documentclass[twoside]{dis07}
\usepackage[latin1]{inputenc}
\usepackage[dvips]{graphicx,epsfig,color}
\usepackage{wrapfig,rotating}
\usepackage{amssymb,amsmath,array}

\def\etajb{\eta^{\rm jet}_{\rm B}}
\def\etaphi{\eta-\varphi}
\def\etjb{E^{\rm jet}_{T,{\rm B}}}
\def\etjet{E_T^{\rm jet}}
\def\etjbj{E^{\rm jet1}_{T,{\rm B}}}
\def\etjbjj{E^{\rm jet2}_{T,{\rm B}}}
\def\etalab{\eta^{\rm jet}_{\rm LAB}}
\def\g2{GeV$^2$}
\def\mz{M_Z}
\def\q2{Q^2}
\def\as{\alpha_s}
\def\asz{\as(\mz)}
\def\asmz#1#2#3#4#5#6{\alpha_s(M_Z) = #1\pm #2\ {\rm (stat.)}\ ^{+#4}_{-#3}\ {\rm (exp.)}\ ^{+#6}_{-#5}\ {\rm (th.)}}
\def\mj{M^{\rm jj}}
\def\m3j{M^{\rm 3j}}

\begin{document}
\title{Summary of $\as$ determinations at ZEUS\footnote{Talk given at
    the ``XV International Workshop on Deep-Inelastic Scattering and
    Related Subjects (DIS07)'', April 16 - 20, 2007, Munich, Germany}}

\author{Claudia Glasman
\thanks{Ram\'on y Cajal Fellow.}
\vspace{.3cm}\\
Universidad Aut\'onoma de Madrid, Spain\\
On behalf of the ZEUS Collaboration
}

\maketitle

\begin{abstract}
The jet cross-section and structure-function measurements done with
the ZEUS detector to extract the strong coupling constant and to test
its energy-scale dependence are summarised. The values of $\as$ thus
obtained and the HERA average are also presented.
\end{abstract}

\section{Introduction}

The strong coupling constant, $\as$, is one of the fundamental
parameters of QCD. However, its value is not predicted by the theory 
and has to be determined experimentally. The success of perturbative
QCD (pQCD) lies on precise and consistent determinations of the
coupling from many diverse phenomena such as $\tau$ decays, event
shapes, $Z$ decays, etc. At ZEUS, many precise determinations of
$\as$ have been performed from a variety of measurements based on
jet observables and on structure functions. 

The procedure to determine $\as$ from jet observables used by
ZEUS is based on the $\as$ dependence of the pQCD calculations and
takes into account the correlation with the proton parton distribution
functions (PDFs). The method consists of performing
next-to-leading-order (NLO) calculations using sets of PDFs for which
different values of $\asz$ were assumed in the fits. A
parameterisation of the $\asz$ dependence of the theory for
the given observable is obtained. Finally, a value for $\asz$
is extracted from the measured cross section using such
parameterisation. This procedure handles correctly the complete
$\as$-dependence of the NLO calculations (the explicit dependence in
the partonic cross section and the implicit dependence from the PDFs)
in the fit, while preserving the correlation between $\as$ and the
PDFs.

\section{Determinations of $\asz$ at ZEUS}

\begin{figure}
\setlength{\unitlength}{1.0cm}
\begin{picture} (20.0,6.5)
\put (0.5,0.0){\epsfig{figure=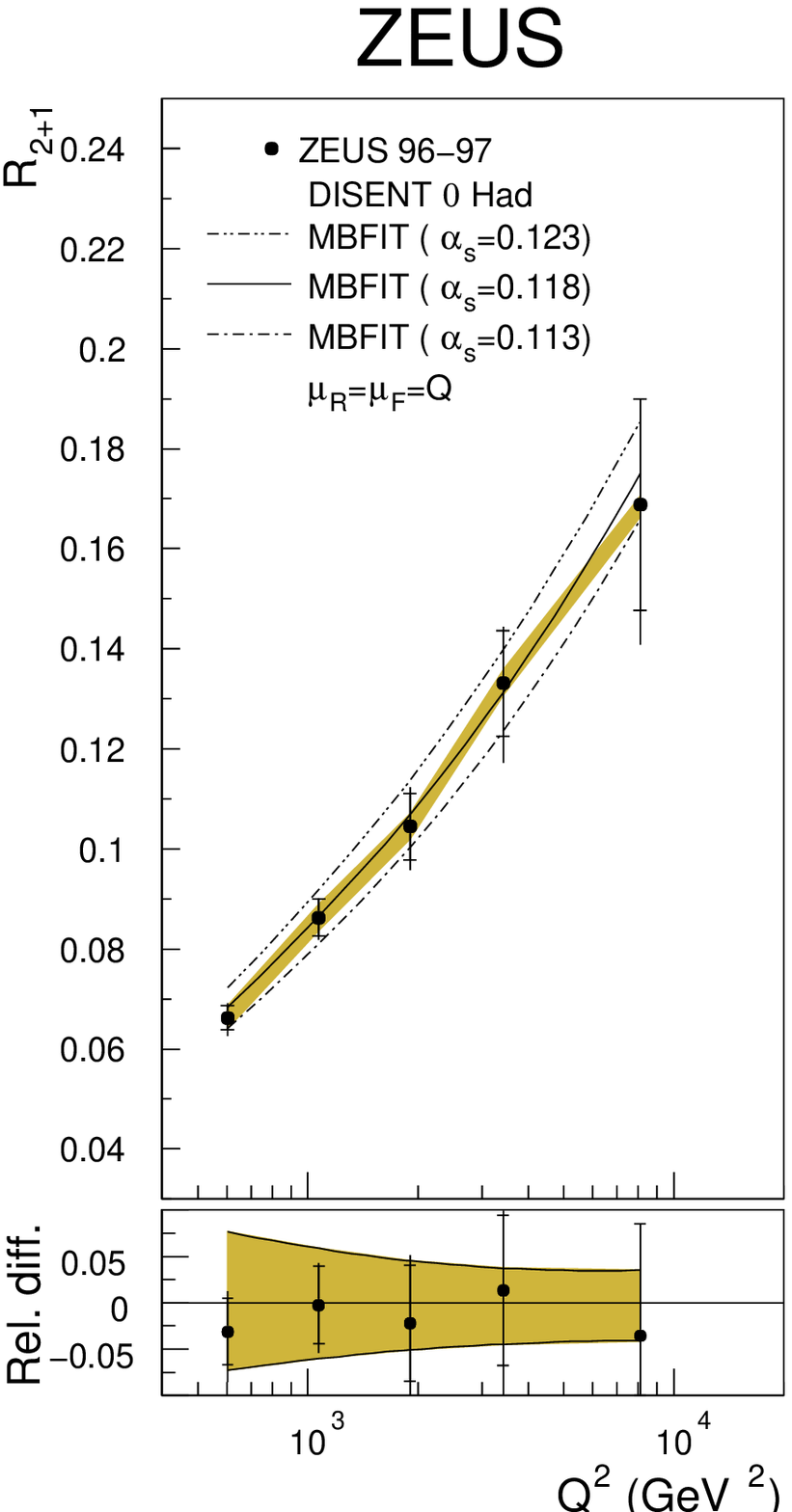,width=3.2cm}}
\put (3.5,0.0){\epsfig{figure=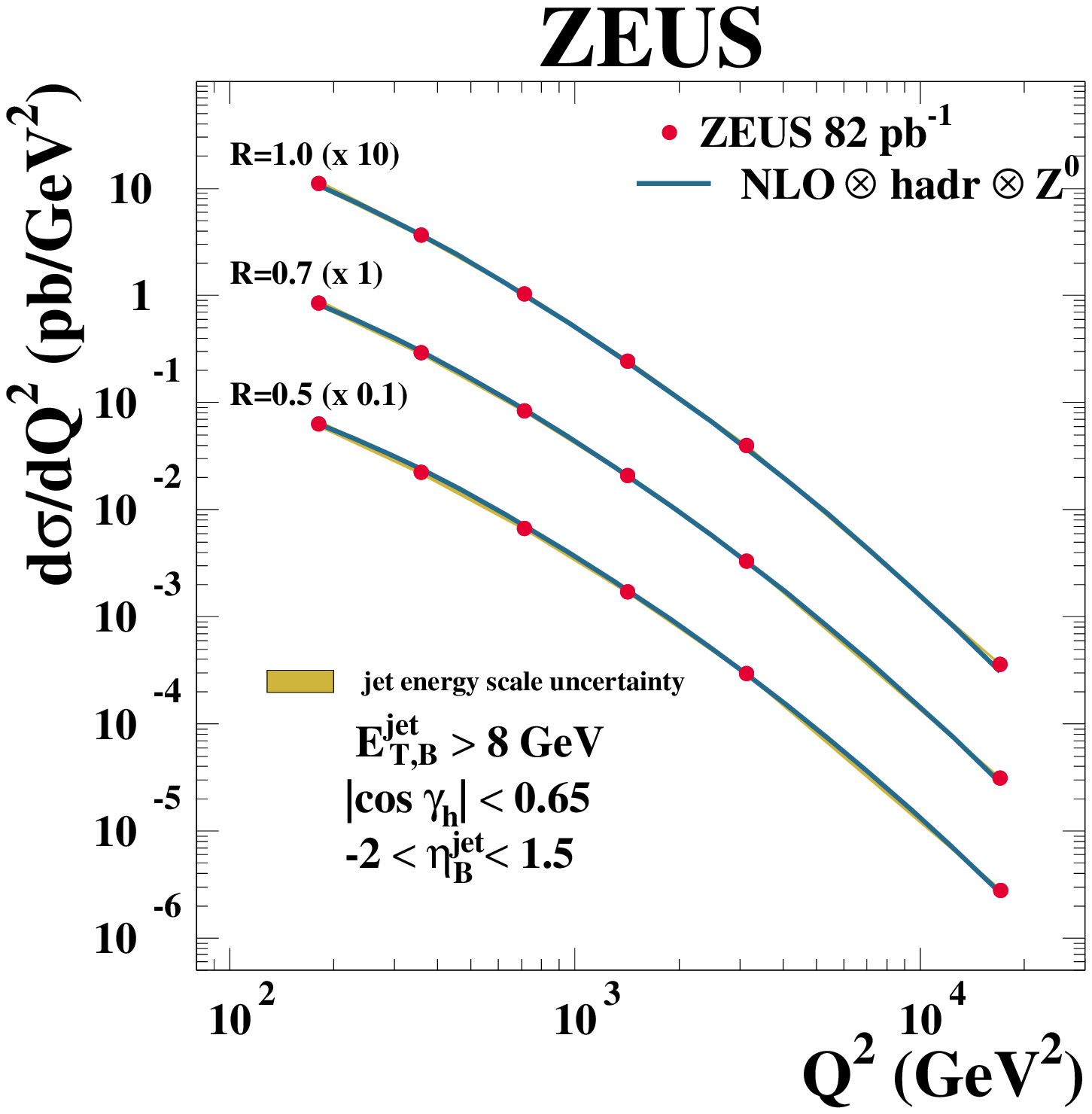,width=6.2cm}}
\put (8.5,0.0){\epsfig{figure=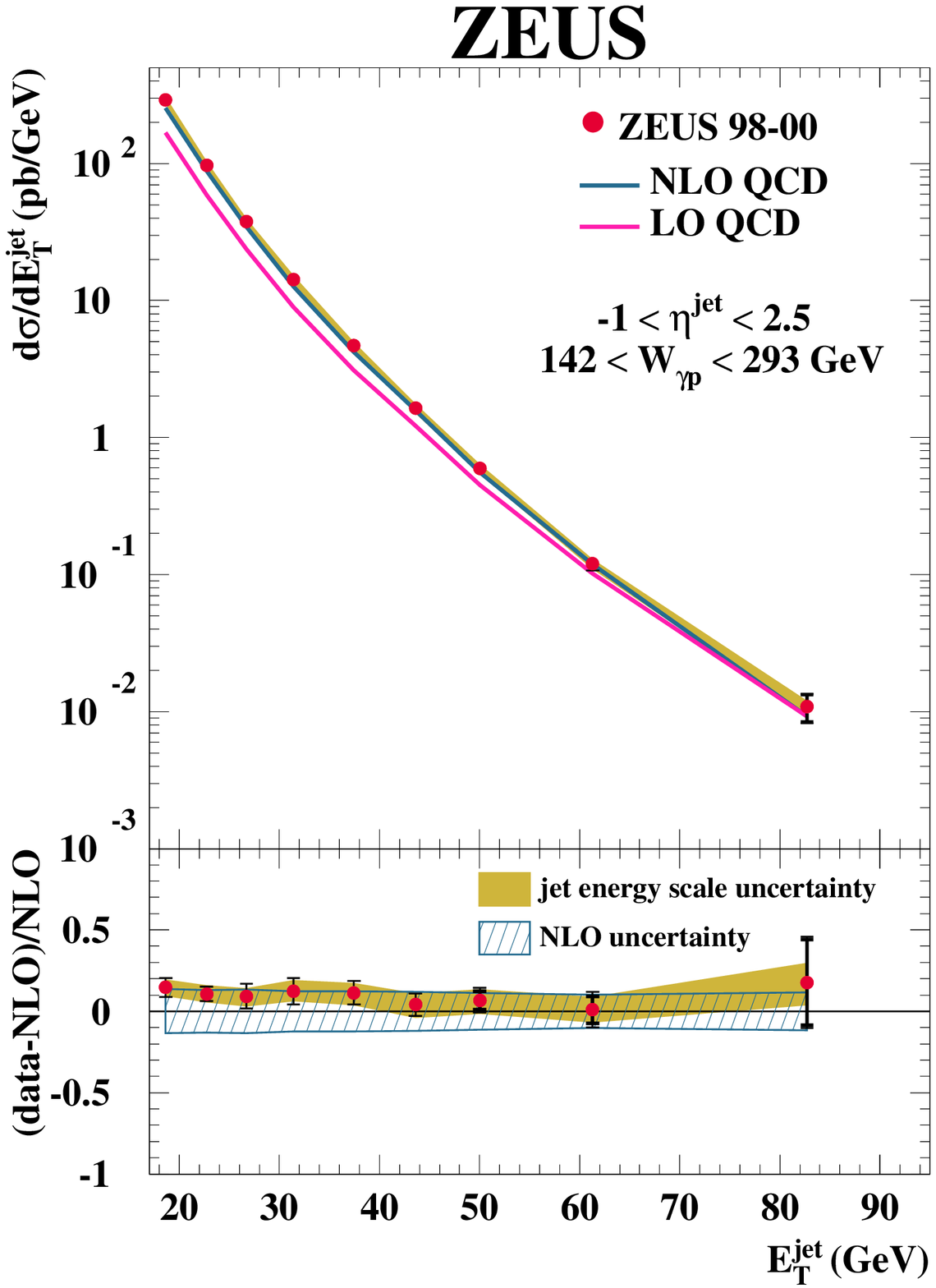,width=6.2cm}}
\put (2.0,-0.3){\bf\small (a)}
\put (6.0,-0.3){\bf\small (b)}
\put (11.5,-0.3){\bf\small (c)}
\end{picture}
\caption{(a) Normalised dijet cross section as a function of $\q2$ in
  NC DIS;
  (b) Inclusive-jet cross section as a function of $\q2$ in NC DIS;
  (c) Inclusive-jet cross section as a function of $\etjet$ in
  photoproduction.
  \label{fig1}}
\end{figure}

The exclusive dijet cross section in neutral-current (NC) deep
inelastic scattering (DIS) has been measured~\cite{dijets}
in the Breit frame in the kinematic region given by
$470<\q2<20000$~\g2, where $\q2$ is the photon virtuality. Two jets
with transverse energies $\etjbj>8$ and $\etjbjj>5$ GeV and
pseudorapidity $-1<\etalab<2$ were selected. Figure~\ref{fig1}a
shows the ratio of the dijet cross section to the total inclusive DIS
cross section as a function of $\q2$. The experimental uncertainties
are small, $\sim 6\%$. The theoretical uncertainties are smaller than
for the individual cross sections. The measured ratio is described
well by the pQCD prediction. The predictions for different values of
$\as$ show the sensitivity of this observable to the coupling. From
the measured ratio for $\q2>470$~\g2, the value
$$\asmz{0.1166}{0.0019}{0.0033}{0.0024}{0.0044}{0.0057},$$
was extracted. In this determination, the theoretical
uncertainties coming from the higher orders dominate.

Inclusive-jet cross sections in NC DIS have been
measured~\cite{inclusive} in the Breit frame in the kinematic region
of $\q2>125$~\g2. Events with at least one jet of $\etjb>8$ GeV and
$-2<\etajb<1.8$ were selected. There are several advantages of
inclusive-jet cross sections with respect to dijet cross sections in a
QCD analysis. The inclusive-jet cross sections are infrared
insensitive; for dijet cross sections asymmetric $\etjb$ cuts are
necessary to avoid the infrared-sensitive regions where the NLO
programs are not reliable. This difficulty is not present in the
calculations of inclusive-jet cross sections so these measurements
allow tests of pQCD in a larger phase-space region than in dijet
production.
Furthermore, the theoretical uncertainties are smaller than
in dijet cross sections. Figure~\ref{fig1}b shows the inclusive-jet
cross section as a function of $\q2$ for different jet radii, $R$. The
measured cross sections are well described by the NLO predictions. The
experimental uncertainties are $\sim 5\%$.
A value of $\as$ has been extracted from the inclusive-jet cross
section with $R=1$ for $\q2>500$ \g2,
$$\asmz{0.1207}{0.0014}{0.0033}{0.0035}{0.0023}{0.0022}.$$
The experimental uncertainties are dominated by the jet
energy scale uncertainty, which amounts to $\pm 2\%$ and the
theoretical uncertainties include the terms beyond NLO ($\pm 1.5\%$),
the uncertainties coming from the proton PDFs ($\pm 0.7\%$) and the
hadronisation corrections ($\pm 0.8\%$). This determination constitutes
the most precise at HERA due to the advantages of using
inclusive-jet cross sections at high $\q2$, with a total theoretical
uncertainty of only $\pm 1.9\%$.

The inclusive-jet cross section in photoproduction has been
measured~\cite{inclusivegp} as a function of $\etjet$ (see
Fig.~\ref{fig1}c). For these processes, transverse energies of up to
$95$ GeV are accessible. The measured cross section shows a steep
fall-off of more than five orders of magnitude within the measured
range. The uncorrelated experimental uncertainties are $\sim 5\%$ at
low $\etjet$ and increase to $\sim 10\%$ at high $\etjet$. The
theoretical uncertainty due to higher orders is $<10\%$, and the
uncertainties due to the parameterisations of the proton and photon
PDFs are $<5\%$. The hadronisation corrections are $\sim 2.5\%$ with
an uncertainty of $2.5\%$. The LO calculation underestimates the data
by $\sim 50\%$ for $\etjet<45$ GeV, whereas the NLO calculation gives
a very good description of the data within the measured range.
The determination of $\as$ from inclusive-jet cross sections in
photoproduction has an additional uncertainty coming from the photon
PDFs, but at the high $\etjet$ covered by the measurements, the
contribution from resolved processes is reduced and so this
uncertainty is of the same order as that coming from the proton
PDFs. Therefore, this determination of $\as$ is also one of the most
precise at HERA,
$$\asmz{0.1224}{0.0001}{0.0019}{0.0022}{0.0042}{0.0054}.$$

%Figure 1

\begin{figure}
\setlength{\unitlength}{1.0cm}
\begin{picture} (20.0,5.0)
\put (-0.2,0.5){\epsfig{figure=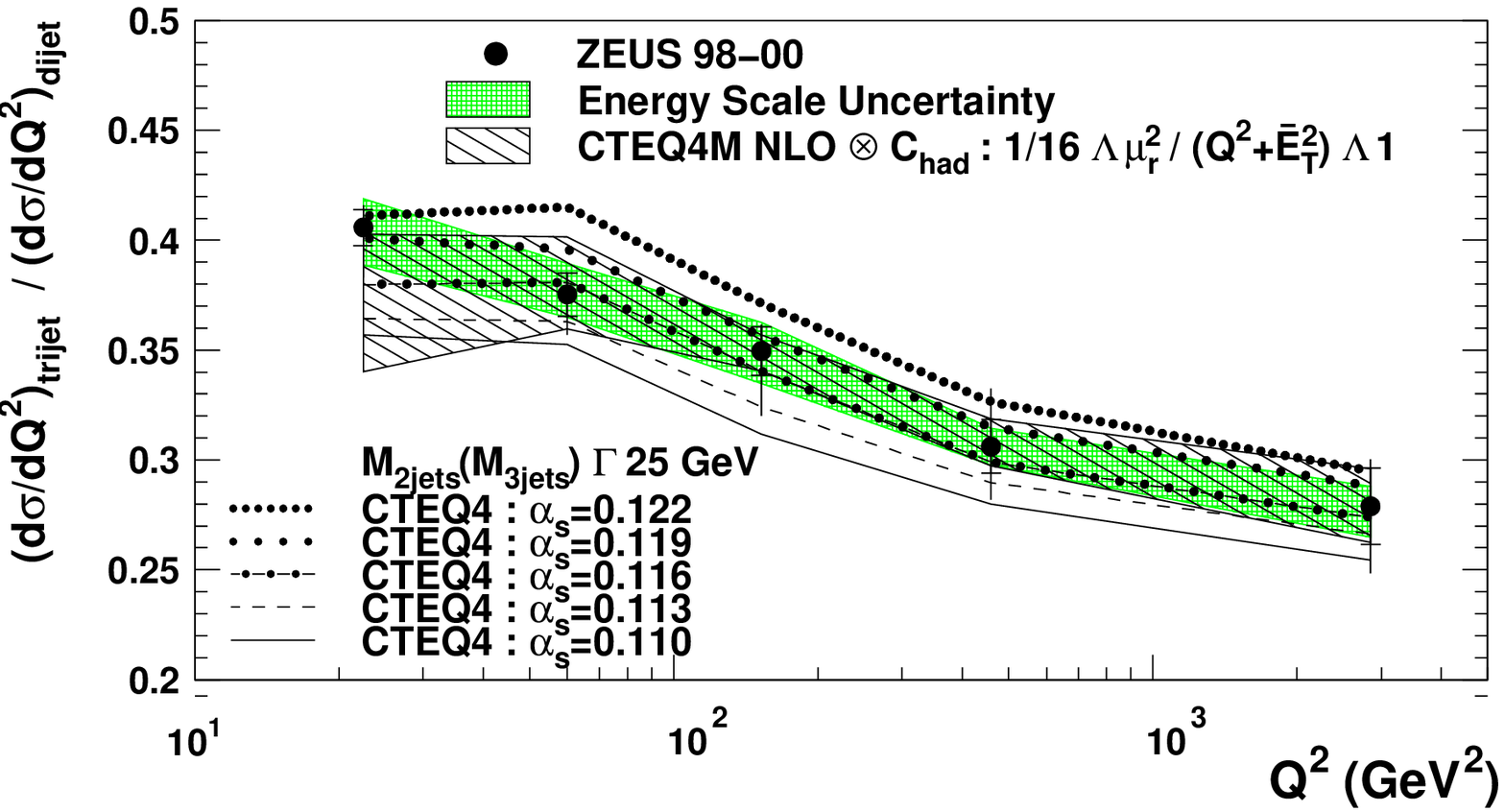,width=6.2cm}}
\put (6.2,0.5){\epsfig{figure=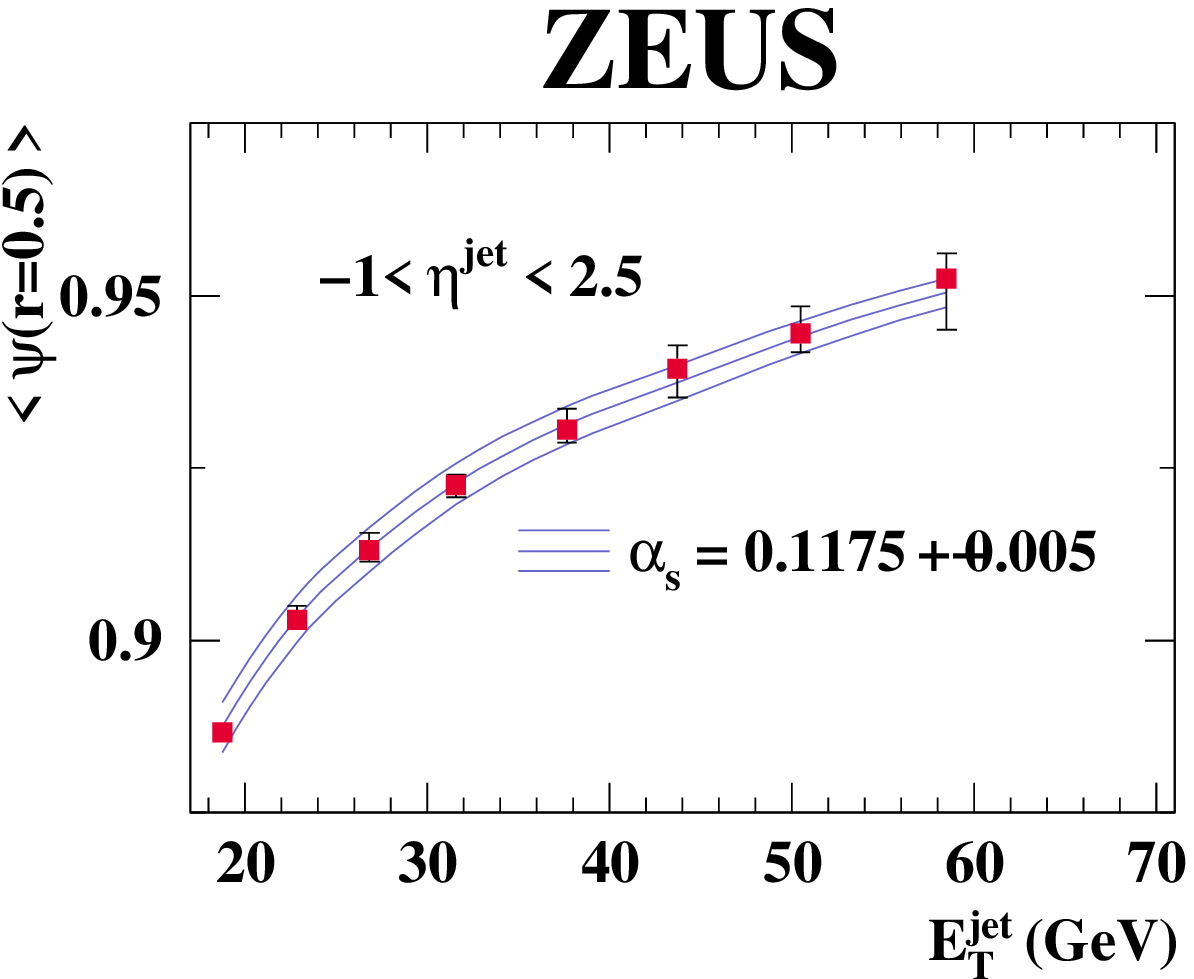,width=4.2cm}}
\put (10.5,-0.2){\epsfig{figure=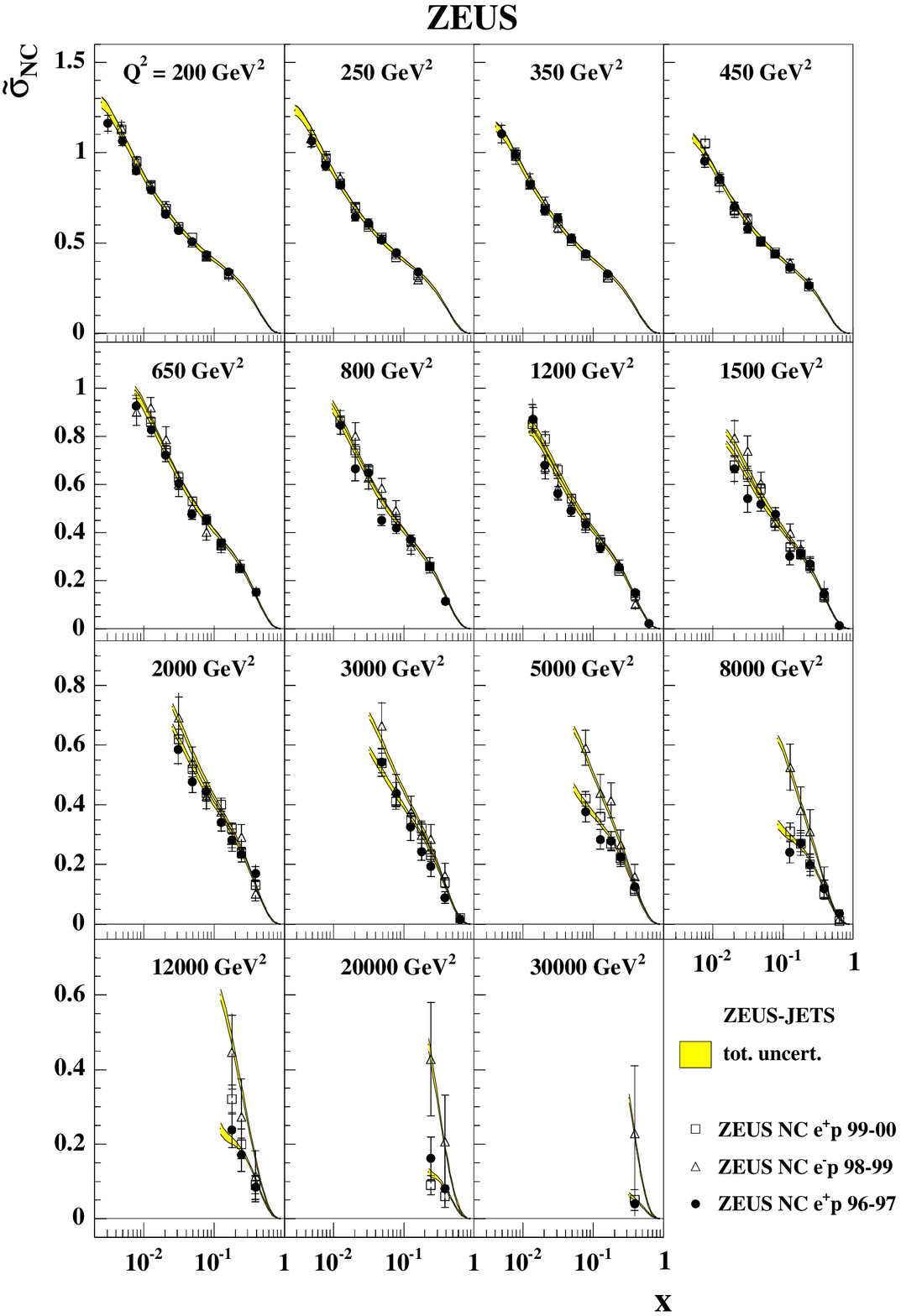,width=3.5cm}}
\put (2.5,-0.3){\bf\small (a)}
\put (8.0,-0.3){\bf\small (b)}
\put (11.5,-0.3){\bf\small (c)}
\end{picture}
\caption{(a) Ratio of trijet to dijet cross sections as a function of
  $\q2$ in NC DIS.
  (b) Mean integrated jet shape as a function of $\etjet$ in NC DIS; 
  (c) Reduced cross section as a function of $x$ in different regions
  of $\q2$ in NC DIS.
  \label{fig2}}
\end{figure}

The dijet (three-jet) cross sections in NC DIS have been
measured~\cite{multijetszeus} as a function of $\q2$ in the Breit
frame for events with at least two (three) jets of $\etjb>5$ GeV and
$-1<\etalab<2.5$, in the kinematic range given by $150<\q2<15000$ \g2\
and $0.2<y<0.6$, where $y$ is the inelasticity. Events with a dijet
(trijet) invariant mass $\mj>25$ ($\m3j>25$) GeV were
selected. Figure~\ref{fig2}a shows the ratio of the trijet to the
dijet cross section as a function of $\q2$. The data are compared to
the predictions of NLO QCD using different values of $\asz$. This
comparison shows the sensitivity of the observable to the value of
$\as$. The measured ratio is described well by the predictions. This
ratio is well suited to determine $\as$ at low $\q2$ since the
correlated experimental and theoretical uncertainties cancel partially
in the ratio. Therefore, this observable provides an accurate test of
color dynamics at low $\q2$, since the theoretical
uncertainty of the ratio is of the same order as at higher $\q2$. From
the measured ratio of trijet to dijet cross sections in the range
$10<\q2<5000$ \g2, a value of $\as$ has been extracted, 
$$\asmz{0.1179}{0.0013}{0.0046}{0.0028}{0.0046}{0.0064},$$
with only $\sim 5\%$ uncertainty coming from higher orders at these
low values of $\q2$.

An independent method to extract $\as$ has been developed which relies
on the detailed description of the internal structure of the jets by
pQCD. The internal structure of jets can be studied by means of the 
integrated jet shape, which is defined as the average fraction of the
jet transverse energy that lies inside a cone in the $\etaphi$ plane
of radius $r$ concentric with the jet axis. The integrated jet shape
has been measured~\cite{jetshape} in NC DIS in the kinematic region
given by $\q2>125$ \g2\ for jets of $\etjet>17$ GeV and
$-1<\etalab<2.5$. Figure~\ref{fig2}b shows the measurements of the 
mean integrated jet shape as a function of $\etjet$ for a fixed value
of $r=0.5$. The measured integrated jet shape increases as $\etjet$
increases. The experimental uncertainties and the corrections for 
detector and hadronisation effects are small for $r=0.5$. The NLO
QCD calculations give a very good description of the data and show the
sensitivity of this observable to the value of $\asz$. The extraction
of $\as$ from the internal structure of jets gives a value with one of
the smallest experimental uncertainties and negligible theoretical
uncertainty coming from the PDFs, but, on the other hand, the
theoretical uncertainty from the higher orders increases to about
$7\%$. The value obtained is
$$\asmz{0.1176}{0.0009}{0.0026}{0.0009}{0.0072}{0.0091}.$$

A fit to inclusive DIS data, such as it is shown in Fig.~\ref{fig2}c,
and jet data has been performed~\cite{nlofitzeusn} to extract
simultaneously the proton PDFs and $\as$. Conventionally, proton PDFs
parameterisations are extracted by fitting inclusive DIS data, which
are directly sensitive to the quark content of the proton; the gluon
density is extracted via scaling violations and sum rules. On the
other hand, jet cross sections are directly sensitive to both the
quark and gluon densities and to $\asz$, with the processes
$\gamma^{(*)}q\rightarrow qg$ not coupled to the gluon
density. Therefore, the inclusion of jet cross sections in the fit
constraints significantly the gluon density and allows an extraction
of $\asz$ from structure functions which is not strongly correlated to
the gluon density. The value obtained from such a fit is
$$\asz=0.1183\pm 0.0028\ ({\rm exp.})\pm 0.0008\ ({\rm model})\pm
0.0050\ ({\rm h.o.}),$$
which yields a very precise determination of $\asz$ from ZEUS data
alone.

%Figure 2

Figure~\ref{fig3}a shows a summary of the values of $\asz$ mentioned
above together with other determinations done at ZEUS. All these
values are in agreement with each other and with the world
average~\cite{bethke}. The experimental uncertainty for the
determinations presented here ranges from $1.8$ to $4.1\%$, whereas
the theoretical uncertainty is between $1.9$ and $7.7\%$. The value with
the lowest theoretical uncertainty is that extracted from the
inclusive-jet cross sections in NC DIS.

\section{An average of $\asz$ at HERA}

To make a proper average of the determinations of $\asz$ from the ZEUS
and H1 Collaborations, the correlations among the different
determinations has to be taken into account. The experimental
contribution to the uncertainty due to that of the energy scale of the
jets, which is the dominant source in the jet measurements, is
correlated among the determinations from each experiment. On the
theoretical side, the uncertainty coming from the proton PDFs is
certainly correlated whereas that coming from the hadronisation
corrections is only partially correlated. The uncertainty coming from
terms beyond NLO is correlated up to a certain, a priori unknown,
degree; since these uncertainties are dominant, special care must be
taken in the treatment of these uncertainties when making an average
of the determinations of $\asz$ at HERA.

A conservative approach has been used to make the
average~\cite{hep-ex/0506035} in which the known correlations among the
determinations of $\as$ coming from the same experiment were taken
into account (``correlation method''). The theoretical uncertainties
arising from terms beyond NLO were assumed to be (conservatively)
fully correlated. Error-weighted averages were obtained separately for
the ZEUS and H1 measurements. Finally, a HERA average was obtained by
using the error-weighted average method on the ZEUS and H1 averages,
assuming the experimental uncertainties to be uncorrelated and taking
the overall theoretical uncertainty as the linear average of its
contribution in each experiment. The average of the HERA
measurements and its uncertainty are~\cite{hep-ex/0506035}:
$$\overline\asz=0.1186\pm 0.0011\ ({\rm exp.})\pm 0.0050\ ({\rm th.}),$$
with an experimental (theoretical) uncertainty of $\sim 0.9\ (4)\%$.
This average, together with the individual values considered, is shown
in Fig.~\ref{fig3}b. It is found to be in good agreement with the
world average (see Fig.~\ref{fig3}a), which does not include
any of these determinations.

\section{Energy-scale dependence of $\as$ at ZEUS and HERA combination}

\begin{figure}
\setlength{\unitlength}{1.0cm}
\begin{picture} (20.0,6.0)
\put (-0.5,-0.3){\epsfig{figure=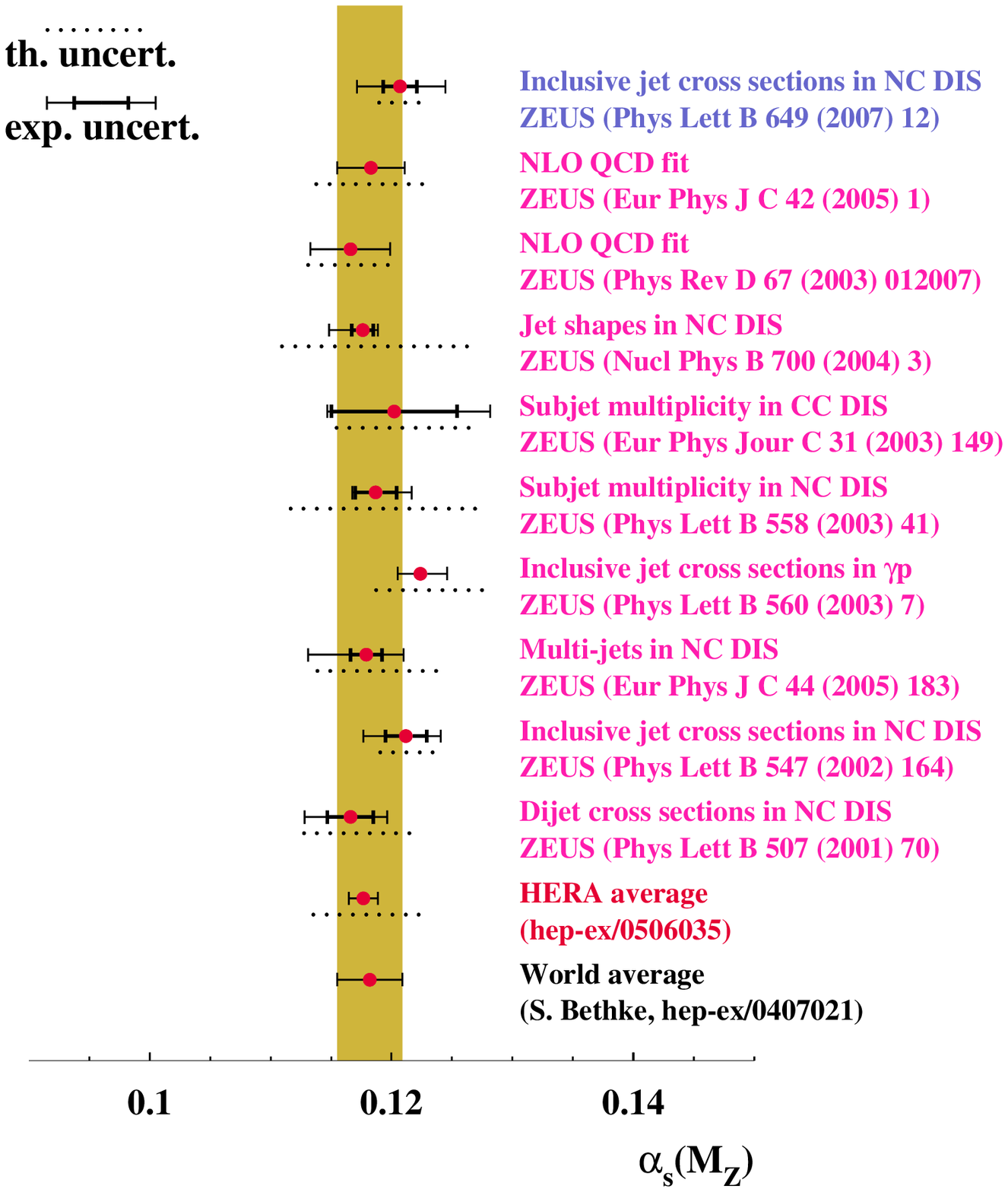,width=6.2cm}}
\put (4.5,-0.5){\epsfig{figure=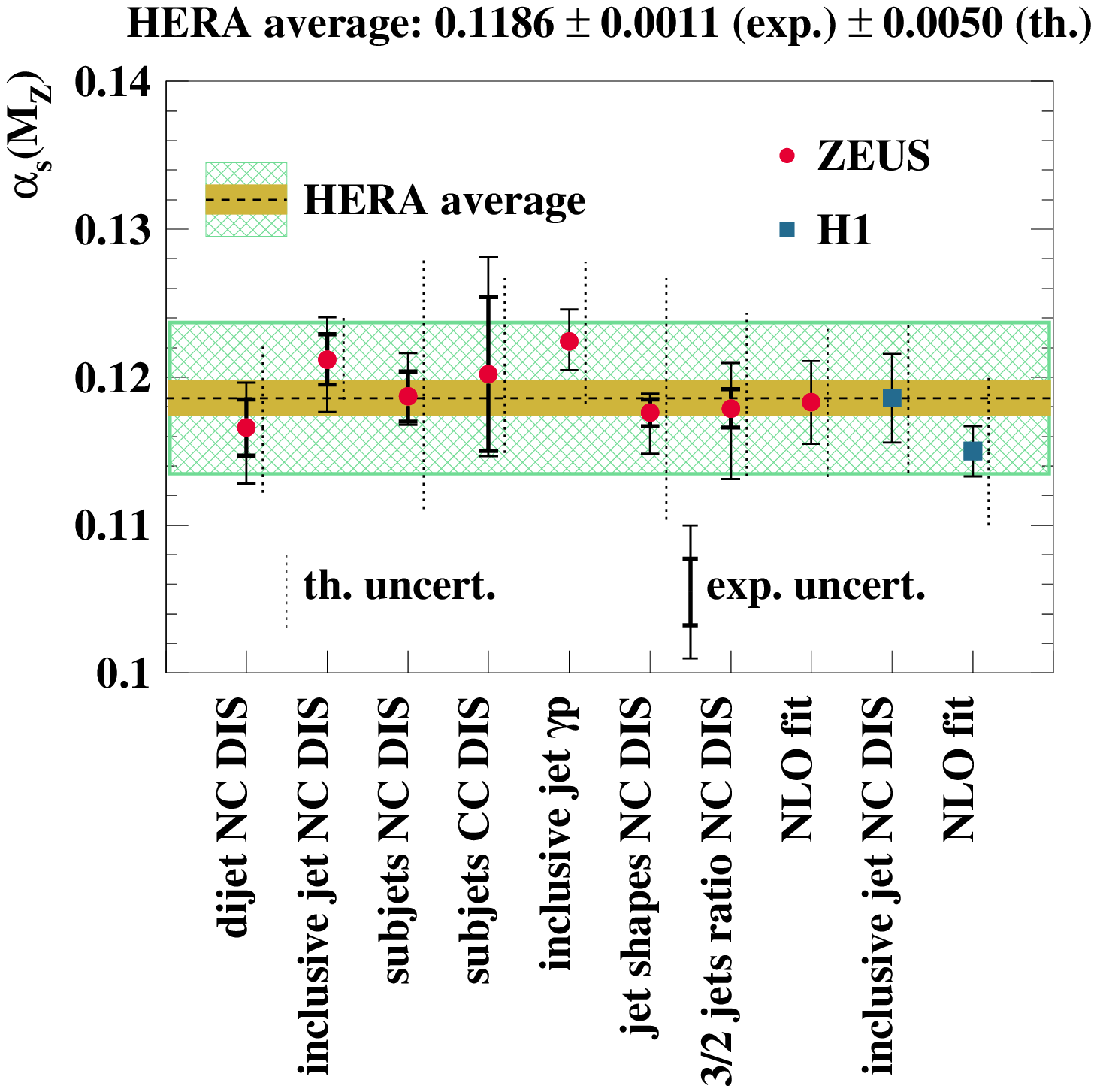,width=6.2cm}}
\put (10.0,2.5){\epsfig{figure=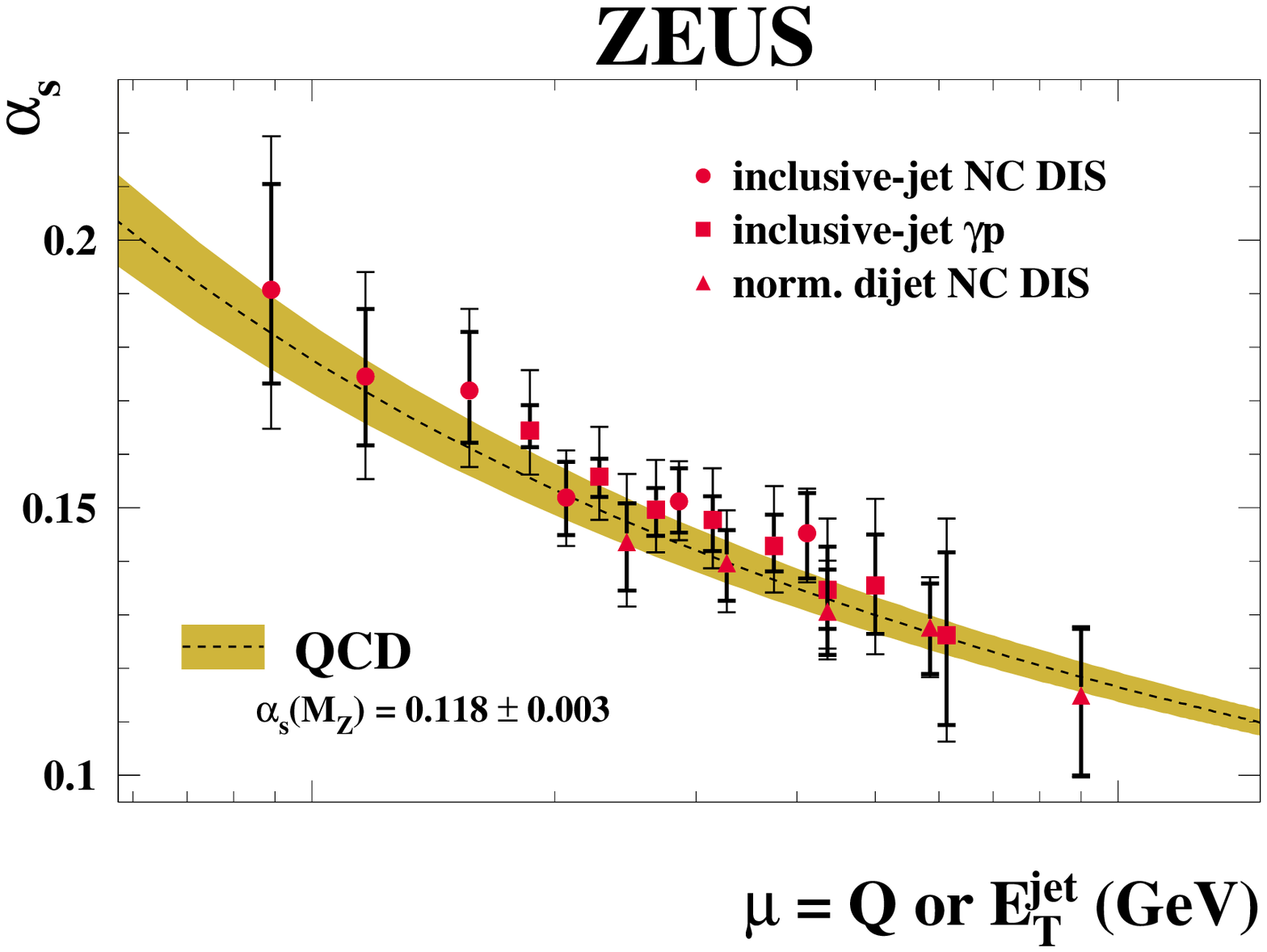,width=4.2cm}}
\put (10.2,-0.5){\epsfig{figure=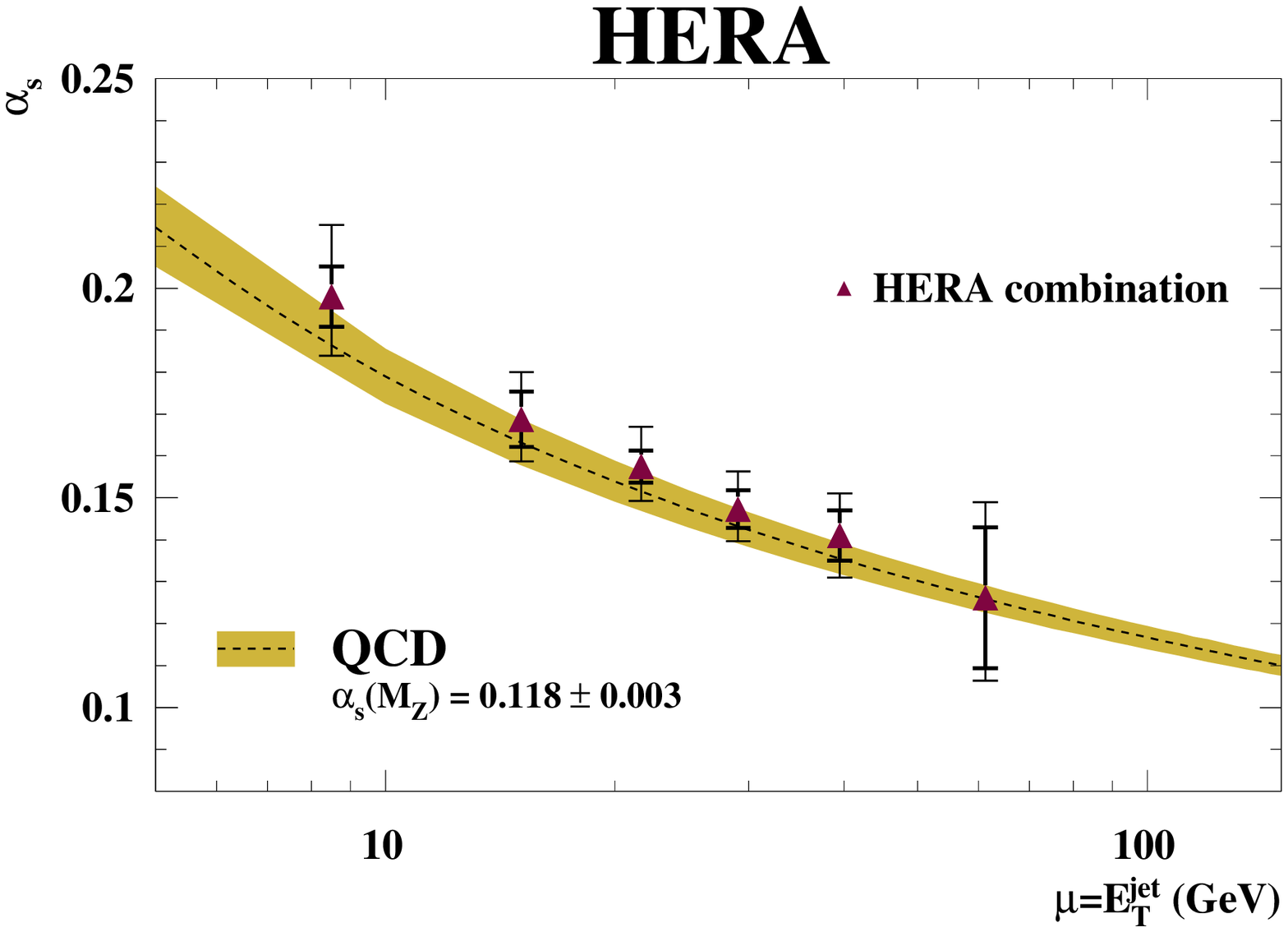,width=4.2cm}}
\put (1.5,-0.3){\bf\small (a)}
\put (7.5,-0.3){\bf\small (b)}
\put (12.0,3.0){\bf\small (c)}
\put (12.0,-0.3){\bf\small (d)}
\end{picture}
\caption{(a) Summary of $\asz$ measurements at ZEUS; (b) HERA $\asz$
  average; (c) Energy-scale dependence of $\as$ at ZEUS; (d)
  HERA-combined energy-scale dependence of $\as$.
  \label{fig3}}
\end{figure}

The ZEUS Collaboration has tested the pQCD prediction of the
energy-scale dependence of the strong coupling constant by determining
$\as$ from the measured differential jet cross sections at different
scales~\cite{dijets,inclusive,inclusivegp}. Figure~\ref{fig3}c shows
the determinations of the energy-scale dependence of $\as$ as a
function of $\etjet$ or $Q$. The determinations are consistent with
the running of $\as$ as predicted by pQCD over a large range in the
scale.

The determinations of $\as(\etjet)$ from the H1 and ZEUS
Collaborations at similar $\etjet$ have been
combined~\cite{hep-ex/0506035} using the correlation method explained
above. The combined HERA determinations of the energy-scale dependence
of $\as$ are shown in Fig.~\ref{fig3}c, in which the running of $\as$
from HERA jet data alone is clearly observed.

%Figure 3

\begin{footnotesize}

\end{footnotesize}


\begin{thebibliography}{99}

\bibitem{dijets}
  ZEUS Collaboration, J. Breitweg et al., Phys. Lett. B 507 (2001) 70.

\bibitem{inclusive}
  ZEUS Collaboration, S. Chekanov et al., Phys. Lett. B 649 (2007) 12.

\bibitem{inclusivegp}
  ZEUS Collaboration, S. Chekanov et al., Phys. Lett. B 560 (2003) 7.

\bibitem{multijetszeus}
  ZEUS Collaboration,  S. Chekanov et al., Eur. Phys. Jour. C 44 (2005) 183.

\bibitem{jetshape}
  ZEUS Collaboration, S. Chekanov et al., Nucl. Phys. B 700 (2004) 3.

\bibitem{nlofitzeusn} 
  ZEUS Collaboration, S. Chekanov et al., Eur. Phys. Jour. C 42 (2005) 1.

\bibitem{bethke}
  S. Bethke, J. Phys. G 26 (2000) R27. Updated in preprint
  hep-ex/0407021.

\bibitem{hep-ex/0506035} 
  C. Glasman, {\em Proc. of the 13th International Workshop on Deep 
    Inelastic Scattering}, S.R. Dasu and W.H. Smith (eds.),
  p. 689. Madison, USA (2005). Also in preprint hep-ex/0506035.

\end{thebibliography}
\end{document}